\documentstyle[aps,multicol]{revtex}

\begin{document}
\draft

\newcommand{\beq}{\begin{equation}} 
\newcommand{\eeq}{\end{equation}}
\newcommand{\bqa}{\begin{eqnarray}} 
\newcommand{\eqa}{\end{eqnarray}}
\newcommand{\nn}{\nonumber} 
\newcommand{\dg}{^\dagger}
\newcommand{\smallfrac}[2]{\mbox{$\frac{#1}{#2}$}}
\newcommand{\bra}[1]{\langle{#1}|} 
\newcommand{\ket}[1]{|{#1}\rangle}
\newcommand{\sch}{Schr\"odinger } 
\newcommand{\schs}{Schr\"odinger's }
\newcommand{\hei}{Heisenberg } 
\newcommand{\heis}{Heisenberg's }
\newcommand{\half}{\smallfrac{1}{2}} 
\newcommand{\bl}{{\bigl(}}
\newcommand{\br}{{\bigr)}} 

\title{Maximally Robust Unravelings of Quantum Master Equations}
\author{H.\ M.\ Wiseman$^{1,2}$ and J.\ A.\ Vaccaro$^{2}$ }
\address{$^{1}$Department of Physics, The University of Queensland, St.\ Lucia 4072, 
Australia \\
$^{2}$Physics Department, The Open University, Milton Keynes
MK7 6AA,  United Kingdom}

\date{\today}
 
\maketitle

\begin{abstract}
The stationary solution $\rho$ of a quantum master equation can be 
represented as an ensemble of pure states in a continuous infinity of 
ways.  An ensemble which is physically realizable through monitoring 
the system's environment we call an `unraveling'.  The survival 
probability $S(t)$ of an unraveling is the average probability for 
each of its elements to be unchanged a time $t$ after cessation of 
monitoring.  The maximally robust unraveling is the one 
for which $S(t)$ remains greater than the largest eigenvalue of $\rho$ for 
the longest time.  The optical parametric 
oscillator is a soluble example.

\end{abstract}   


\begin{multicols}{2}

It is well-known that open quantum systems  
generally become entangled with their environment. Thus they do 
not remain pure, but become mixed. Nevertheless, pure states are 
used in the analysis of open quantum systems, in at least two distinct 
contexts. The first context is that of decoherence (see 
for example Refs.~\cite{Gea90,ZurHabPaz93,BarBurVac96}). The pure states 
which are studied in this field are those which are relatively 
long-lived (or decohere slowly) under a particular 
irreversible evolution. In Ref.~\cite{ZurHabPaz93} they are called 
{\em stable states}. The 
second context is that of continuous quantum measurement theory
\cite{Car93b,QSOSQO96}. Under some situations, 
the perfect monitoring of the system's environment collapses the system to 
a pure state which undergoes a stochastic evolution, called a quantum 
trajectory in Ref.~\cite{Car93b}. 
The ensemble of quantum trajectories 
reproduces the deterministic dissipative evolution, and so is called an {\em 
unraveling} of that evolution.

In this letter we unite these two threads in the idea of a 
{\em maximally robust unraveling}. We consider systems obeying a quantum 
master equation, which is the most general generator of 
deterministic Markovian evolution and  
is used in many areas of physics. We further assume the system
to have a unique stationary state matrix $\rho_{\infty}$. Monitoring 
the bath to which the system is coupled will give unravelings
which are generated by a (nonlinear) stochastic \sch equation (SSE).
Different ways of monitoring the bath have different SSEs, but each 
SSE has a stationary ensemble of pure state solutions which
represents $\rho_{\infty}$. Here we calculate the 
{\em average} survival probability $S(t)$ of these {\em physically realizable 
ensembles}, rather than the survival probability
 of a single arbitrarily chosen pure state. 
 As we explain below, 
the maximally robust unraveling is that for which $S(t)$ 
remains higher than the largest eigenvalue of 
$\rho_{\infty}$ for the longest time. 
As an example we consider continuous 
Markovian unravelings of an optical parametric oscillator.

{\em Unraveling the master equation.}---
The most general form of the quantum master equation is 
\cite{RigGis96,Wis96a}
\beq
\dot{\rho}= -i[H,\rho] + \sum_{k=1}^K c_k{\rho}c_k\dg - \half\{c\dg_k c_k 
,\rho\} \equiv 
{\cal L}\rho \label{genme}
\eeq 
where $\{a,b\}\equiv ab+ba$. The 
unique stationary state is defined by 
${\cal L}\rho_\infty = 0$. This requires that ${\cal L}$ be 
time-independent, which implies that 
one can only remove a Hamiltonian term by
moving into its interaction picture if the remainder of the 
superoperator ${\cal L}$ is time-independent in that interaction picture. 

It is
now well-known that there are many (in fact 
continuously many) different unravelings for a given master equation 
\cite{QSOSQO96}. If we represent the pure state of the system by the projector
$P$, we can write any SSE unraveling the master equation (\ref{genme}) as 
\beq \label{SSE1}
d{P}=  dt\left( {\cal L} + {\cal U} \right)P.
\eeq %[\{c_k\}]
Here ${\cal U}$ (which depends on the operators $c_k$) is a
stochastic superoperator which is, in general, nonlinear in its operation on $P$. It
is constrained by the following two equations which must hold for arbitrary
projectors $P$ 
\bqa
 [({\cal L} + {\cal U})P ]P
+ dt [{\cal U}P ][{\cal U}P ] &=& (1-P)[({\cal L} + {\cal U})P] \label{prop1}\\
{\rm E}[{\cal U}P]&=& 0.
\eqa
The former property ensures that $P+dP$ remains a projector and the latter that 
\beq
d {\rm E}[P] = {\cal L}{\rm E}[P] dt, \label{ensav}
\eeq
where ${\rm E}$ denotes the ensemble average with respect to the stochasticity of
${\cal U}$. This stochasticity is evidenced by the necessity of 
retaining the term $dt [{\cal U}P][{\cal U}P ]$ in Eq.~(\ref{prop1}).

For simplicity we will call ${\cal U}$ an unraveling. 
Each unraveling gives rise to an ensemble of pure
states   \beq \label{ens}
E^{\cal U} = \{ P_i^{\cal U}, w_i^{\cal U} \},
\eeq
where $P_i$ are the possible pure states of the system at steady state,
and $w_i$ are their weights. For master equations with a unique
stationary state $\rho_{\infty}$, 
%the SSE (\ref{SSE1}) is ergodic over $E^{\cal U}$ and  
$w_i$ is
equal to the proportion of time the system spends in state $P_i$.
The ensemble $E^{\cal U}$ represents $\rho_{\infty}$ in that 
\beq \label{decomp}
\sum w_i^{\cal U} P_i^{\cal U} = \rho_{\infty},
\eeq
as guaranteed by Eq.~(\ref{ensav}). Note that the projectors $P_i^{\cal U}$ 
will not in general be orthogonal.  

{\em Robustness.}--- Let the system evolve under an unraveling 
${\cal U}$ 
from an initial state at time $-\infty$ 
to the stationary ensemble at time $0$. 
 It will then be in the state $P_i^{\cal U}$ with probability
$w_i^{\cal U}$. If we cease to monitor the system at this time $0$ then  the state
will no longer remain pure, but rather will relax toward  $\rho_{\infty}$ 
under the evolution of Eq.~\ref{genme}). The {\em
robustness} of a particular state $P_i^{\cal U}$ is measured by its survival
probability $S_i^{\cal U}(t)$. This is the probability that it would be found  
(by a hypothetical projective measurement) to still be in the state 
$P_i^{\cal U}$ at time $t$ 
\beq
S_i^{\cal U}(t) = {\rm Tr}[P_i^{\cal U} \exp({\cal L}t)P_i^{\cal U}].
\eeq
The average survival probability of the ensemble $E^{\cal U}$ resulting from the
unraveling ${\cal U}$ is therefore 
\beq
S^{\cal U}(t) = \sum w_i^{\cal U} S_i^{\cal U}(t).
\eeq

In the limit $t\to\infty$ the ensemble average survival probability will tend
towards the stationary value  $S^{\cal U}(\infty) = {\rm Tr}[ \rho_{\infty} ^2]$
which is independent of the unraveling ${\cal U}$ and is a measure of 
the mixedness of $\rho_{\infty}$. An alternate measure of the mixednes of 
$\rho_\infty$ is its largest eigenvalue
\beq
\Lambda = \lim_{n\to \infty}\left({\rm Tr}[\rho_{\infty}^{n}]\right)^{1/n}
= {\rm max}\{\lambda \,:\, \rho_{\infty}Q_{\lambda} = \lambda 
Q_{\lambda} \} .
\eeq
In the absence of any monitoring of the bath, the projector $Q_{\Lambda}$ 
would be one's best guess for what pure state the system is 
in at steady state. The chance of this guess being correct is 
simply $\Lambda$ and is obviously independent of time. Since 
$S^{\cal U}(0)=1\geq \Lambda \geq S(\infty)$, there will be some 
time $t'$ at which the ensemble average survival probability $S^{\cal 
U}(t')$ becomes less than or equal to $\Lambda$. That is, 
given that we know by monitoring the environment the state 
$P^{\cal U}_{i}$ which is occupied at 
time $0$ then {\em at time $t'$ this state ceases (on average) to be any 
better than $Q_{\Lambda}$ as an estimate of which pure 
state is occupied by the system.}  In effect, the ensemble $E^{\cal U}$ 
is obsolete at time $t'$, so it is natural
 to define the {\em survival time} of $E^{\cal U}$ as
\beq \label{deftau}
\tau^{\cal U} = {\rm min} \{ t\,:\, S^{\cal U}(t)= \Lambda  \}.
\eeq
 Note that this expression depends on the entire superoperator ${\cal 
L}$, and so is sensitive to the interplay of Hamiltonian and 
irreversible terms. This is as opposed to the reciprocal rate of decay 
of $S^{\cal U}(t)$ at $t=0$ which depends only on the non-Hamiltonian terms in
${\cal L}$ \cite{BarBurVac96}.  
%, which indicates that the members of the ensemble $E^{\cal U}$  
%are relatively unaffected the action of $\exp({\cal L}t)$.

The survival time $\tau^{\cal U}$ quantifies the robustness of an 
unraveling ${\cal U}$.  Let the set of all unravelings be denoted $J$.
Then the set of {\em maximally robust} unravelings $J_{M}$ is 
\beq
J_{M} = \{ {\cal R} \in J : \tau^{\cal R} \geq \tau^{\cal U} \; \forall \, {\cal 
U} \in J \}.
\eeq
Even if $J_{M}$ has many elements ${\cal R}_{1}, {\cal R}_{2}, \ldots$, 
these different unravelings may 
give the same ensemble $E^{\cal R}=E^{{\cal R}_{1}}=E^{{\cal R}_{2}}=\ldots$. 
There is clearly a strong reason to regard $E^{\cal R}$  
 as the most natural ensemble representation of the
stationary solution of a given master equation. 

{\em Continuous Markovian unravelings.}---
The formalism we have presented is well-defined but is difficult to apply 
because of the size of the set $J$ of allowed unravelings. 
Although the
stochasticity in the superoperators ${\cal U}$ can always be written in terms of
quantum jumps, these jumps range in size from being infinitesimal, 
to being so large that the system state after the jump is always
orthogonal to that before the jump \cite{RigGis96}. 
Also the unravelling need not be Markovian,
even though the master equation is \cite{Wis96a}.  
%That is, ${\cal U}$ at time $t$ can depend
%upon the results of the monitoring of the outputs of the system at times earlier
%than $t$ \cite{Wis95c}
For this reason it is useful to consider a smaller (but still continuously
infinite) set $J'$ containing only continuous Markovian unravelings.
A continuous (but not differentiable) time evolution arises from infinitely  
small (and infinitely frequent) jumps. In this case 
the probability distribution for the pure states obeying the SSE
satisfies a Fokker-Planck equation \cite{Gar85}. These unravelings are thus the natural ones
to consider for quantum systems expected to show quasi-classical 
behaviour \cite{RigGis96}.
%, since the Fokker-Planck equation is the usual classical analogue 
%for a quantum master equation.  

The elements ${\cal U}'$ of $J'$ can be written as  
\beq
{\cal U}' dt = \sum_{k=1}^K {\cal H}[dW_k(t) c_k].
\eeq%[\{c_k\}]
Here ${\cal H}[r]$ is a nonlinear superoperator defined (for an arbitrary 
operator $r$) by
\beq
{\cal H}[r]{\rho} \equiv r\rho + \rho r\dg - {\rm Tr}[r\rho + \rho r\dg]\rho
\eeq
and the $dW_k(t)$ are the infinitesimal increments of a complex 
multi-dimensional Wiener process \cite{Gar85} satisfying
\beq
dW_j(t)dW_k^*(t) = dt \delta_{jk}, \;\;\;dW_j(t)dW_k(t) = dt u_{jk} ,
\eeq
where the $u_{jk}$ are arbitrary complex numbers obeying
$|u_{jk}| \leq 1$ and $u_{jk}=u_{kj}$.
%For a master equation with $K$ Lindblad terms  
%the problem of finding
%the maximally robust unravelling ${\cal R}' \in M'$ thus reduces to a search over
%the bounded region $\{ u_{jk} \,:\, |u_{jk}|^2 \leq 1\}$ in
%$K(K+1)$-dimensional  Euclidean space. 

{\em An example.}---
As an application for our concept of robust unravelings we consider 
the master equation  
\beq
{\cal L}\rho = -({\chi}/{4})[a^2-(a\dg)^2,\rho]+a\rho a\dg - 
\half\{a\dg a,\rho\} . \label{me3}
\eeq
Here $a$ is the annihilation operator for a harmonic oscillator, satisfying
$[a,a\dg]=1$. For $|\chi| < 1$ this master equation is a good model for a
degenerate optical parametric oscillator below threshold \cite{Car93b}. 
Time is being measured
in units of the cavity lifetime. The free Hamiltonian $H_0 = \omega a\dg a$ is
omitted because we are working in its interaction picture.
This is necessary for the Hamiltonian term in Eq.~(\ref{me3}), 
which arises from driving a
$\chi^{(2)}$ crystal with a classical field at frequency $2\omega$, to be
time-independent. The damping term remains time-independent in the interaction
picture. 

 For $|\chi| < 1$ the equation ${\cal L}\rho_\infty=0$ has a solution, 
 and it is a Gaussian state. 
 That is to say, the symmetrized moments for the quadrature
operators $x = a+a\dg$ and $y= -ia+ia\dg$  are those of a bivariate Gaussian
distribution (which is in fact the Wigner function \cite{Moy49} for the state). Such states are
characterized by a mean vector and a covariance matrix which we denote 
\bqa
\vec{\mu} &=& (\langle x \rangle , \langle y \rangle)^T = (\bar{x},\bar{y})^T ,\\
{\bf M} &=& \left( \begin{array}{cc} \langle x^2 \rangle & 
\half\langle xy+yx \rangle \\
   \half\langle xy+yx \rangle & \langle y^2 \rangle 
                   \end{array} \right) = 
\left( \begin{array}{cc} \gamma & \beta \\
   \beta & \alpha 
                   \end{array} \right).
\eqa
For all Gaussian quantum states the second-order moments obey the following
inequality
\beq
\left({\rm Tr}[\rho^{2}]\right)^{-2}=\det({\bf M}) = \alpha\gamma - \beta^2 \geq 1, \label{bup}
\eeq
where the equality holds only for pure states.
 For $\rho_{\infty}$ we find 
 ${\bf M}_{\infty} = {\rm diag}[(1+\chi)^{-1},(1-\chi)^{-1}]$ and 
 $\vec\mu_{\infty}=\vec{0}$.
The maximum eigenvalue for this state is
\beq \label{maxev}
\Lambda = 
%2\left( 1+\sqrt{\det{\bf M}_{\infty}}\right)^{-1} = 
2\left[ 1+(1-\chi^{2})^{-1/2}\right]^{-1}.
\eeq
 
 The continuous Markovian SSEs which unravel the master equation (\ref{me3}) are of
the form  \beq \label{SSE2}
dP={\cal L}P dt + {\cal H}[dW(t)a]P,
\eeq
where $dW^*(t)dW(t)=dt$. Because there is
only one irreversible term in the master equation, these unravelings are
parameterized by the single complex number $u$ satisfying $|u|\leq 1$ and defined
by  
\beq 
dW(t)^2 = udt = (r+ih)dt.  \label{corfn}
\eeq
The different unravelings (hereafter denoted by $u$) 
could be physically realized by dividing the cavity
output into two beams so that two different quadratures may be 
measured simultaneously by homodyne
detection.

To find the ensemble $E^u$ we need to solve the SSE
(\ref{SSE2}). This can be done
by assuming a Gaussian (as defined above) initial state. Then, 
following the method of 
Ref.\cite{Wis93a}, we can show that this state will remain Gaussian, 
with its moments obeying the differential equations
\bqa
d{\bar{x}} &=& -(1+\chi)(\bar{x}/2){dt} + {\rm Re} \left\{
{dW(t)}(\gamma - 1 -i\alpha) \right\}, \label{first} \\
d{\bar{y}} &=& -(1-\chi)(\bar{y}/2){dt} + {\rm Re} \left\{
{dW(t)}( \beta -i\alpha + i ) \right\}, \label{second}\\
\dot{\gamma} &=& -(1+\chi)\gamma + 1  + (1/2)\left\{ -({1+r})(\gamma-1)^2  
\right. \nn \\
&& -\, \left. ({1-r}) \beta^2 + 2h(\gamma-1)\beta  \right\} ,
\label{shoe} \\
\dot{\alpha} &=& -(1-\chi)\alpha + 1 + (1/2)\left\{ -({1-r})(\alpha-1)^2 \right.
\nn \\ && -\, \left.  ({1+r}) \beta^2 + 2h(\alpha-1)\beta  \right\}, \\
\dot{\beta} &=& -\beta  + (1/2)\left\{ -({1+r})(\gamma-1)\beta - 
  ({1-r}) (\alpha-1)\beta \right. \nn \\
&& +\, \left. {h}[\beta^2+(\gamma-1)(\alpha-1)] \right\} . \label{last} \eqa
Here the terms arising from the unraveling ${\cal U}={\cal H}[dW(t)a]$ 
have been enclosed in curly brackets.  

 What is remarkable about these equations is that the second-order moments evolve
deterministically. It can be verified numerically that they have a
stationary solution ${\bf M}^{u}_{0}$ which satisfies the equality in 
Eq.~(\ref{bup}), as expected. 
%This is true even if one starts with a mixed initial state with 
%$\det({\bf M}) > 1$. 
For a given unraveling $u=r+ih$, the system will evolve from an
arbitrary state at time $-\infty$ into a unique ensemble at time $0$ consisting of
Gaussian pure states $P_{\vec\mu}^{u}$, all with the same second-order 
moments ${\bf M}^{u}_{0}$ but distinguished by their first-order moments $\vec{\mu}$.
 Thevector $\vec\mu$ thus acts as the index
$i$ for the ensemble as in Eq.~(\ref{ens}). 
Furthermore, because ${\bf M}^{u}_{0}$ is
constant, Eqs.~(\ref{first}) and (\ref{second}) 
for the first-order moments are those of a two-dimensional Ornstein-Uhlenbeck
equation \cite{Gar85}. 
The stationary probability distribution $w_{\vec\mu}$ for $\vec{\mu}$ is
therefore  Gaussian. By virtue of Eq.~(\ref{ensav}) the 
second-order moments of this distribution plus those of the 
pure states $P_{\vec{\mu}}^u$ must 
equal those of $\rho_{\infty}$. That is, $w^u_{\vec{\mu}}$ is a 
bivariate Gaussian in $\vec\mu$ with mean
$\vec{0}$ and covariance matrix ${\bf M}_{\infty}-{\bf M}^u_0$. 

 Due to the nonlinearity of the equations of motion for ${\bf M}^{u}$, there is no
analytical expression for the stationary solutions ${\bf M}^u_0$ in terms
of $u$. However the inverse problem is easy to solve because the three 
moment equations (\ref{shoe})--(\ref{last}) with the left-hand sides 
equal to zero  are linear in $r$ and $h$. That is,
given the moments ${\bf M}_0^u$ satisfying $\det({\bf M}_0^u)=1$, these three
equations are consistent and yield values of $r$ and $h$ in terms of 
$\gamma,\alpha,\beta$. Of course only those
solutions satisfying $r^2 + h^2 \leq 1$ correspond to physically realizable
unravelings. This easily yields the equation for the set of
allowed moments ${\bf M}_0^{u}$. This region is plotted in Fig.~1 for $\chi=0.9$. 
Only
two axes (namely $\beta^{u}_0,\gamma^{u}_0$) are needed because the equation 
$\det({\bf M}_0^{u})=1$ implies $\alpha^u_0=[(\beta^u_0)^2+1]/\gamma_0^u$. 

This plot of the
physically realizable ensembles contains information about the system not
revealed by previous work.
In particular the set of these ensembles is smaller than the set of all ensembles
of Gaussian pure states  with fixed second-order moments but varying first-order
moments which reproduce the stationary state $\rho_{\infty}$.
 The latter set of second-order moments ${\bf M}$  
 is defined by the two equations 
 $\det({\bf M})=1$ and $\det({\bf M}_{\infty}-{\bf M}) \geq 0$. This 
 region is also plotted in Fig.~1.

%In fact as $\chi \to 1$ the latter set becomes unbounded (allowing all values of
%$\mu_{11}$, while the set of physically realizable ensembles remains bounded. 

%In summary each unravelling $u$ is associated with an ensemble of Gaussian
%projectors $P^u_{\vec{\mu}}$ all with the same second-order moments ${\bf M}^u_0$.
To find the survival probability we need the time-evolved state  
$e^{{\cal L}t}P^u_{\vec{\mu}}$, which is still Gaussian. From
Eqs.~(\ref{first})--(\ref{last}) with the terms in curly brackets removed its
moments are found to be
\beq
\vec{\mu}_t = {\bf V}_t \vec{\mu}, \;\;\;
{\bf M}^{u}_t = {\bf V}_t {\bf M}_0^{u} {\bf V}_t^T + {\bf M}_\infty - 
{\bf V}_t {\bf M}_\infty {\bf V}_t^T ,
\eeq
where ${\bf V}_t={\rm diag}(v_{+},v_{-})$, where $v_\pm=\exp\bl-(1\pm \chi)t/2\br.$
%For $t \to \infty$ we have $\vec{\mu}_\infty = \vec{0}$ and ${\bf M}_\infty={\bf
%M}_\infty$.  
Using the above results and the properties of Wigner functions 
\cite{Moy49} we find
\end{multicols}
\vspace{-0.5cm}
\noindent\rule{0.5\textwidth}{0.4pt}\rule{0.4pt}{\baselineskip}
\widetext
\bqa
S^u(t) &=& \int d^2\vec\mu\; w^u_{\vec{\mu}}\; {\rm Tr}[P^u_{\vec{\mu}} 
\exp({\cal L}t)P^u_{\vec{\mu}}] 
\,=\, \left\{ \det\bl \half [ (1-{\bf V}_t)({\bf M}_\infty-{\bf
M}_0^{u})(1-{\bf V}_t^T) + {\bf M}_t^{u} + {\bf M}_0^{u} ] \br \right\}^{-1/2} , \\
%\,=\, \left\{ \det\bl {\bf M}_\infty - \half{\bf V}_t({\bf M}_\infty-{\bf M}^{u}_0) -
%\half ({\bf M}_\infty-{\bf M}^{u}_0){\bf V}_t^T \br \right\}^{-1/2}. 
%\\
%\eqa
%An equation of this form is true of Gaussian systems
%in arbitrary dimensions. For the case at hand it evaluates to
%\bqa
%S^u(t)
 &=& \left\{ v_+v_- + \frac{v_-(1-v_+)}{(1+\chi)\gamma_0^u} +
\frac{v_+(1-v_-)\gamma_0^u}{1-\chi}+\frac{(1-v_-)(1-v_+)}{1-\chi^2}  +
(\beta^u_0)^2\left[ \frac{v_-(1-v_+)}{(1+\chi)\gamma^u_0} -
\left(\frac{v_+-v_-}{2}\right)^2 \right] \right\}^{-1/2} \!\!\!\!.
\eqa
\begin{multicols}{2}

It is easy to prove
that for $0<\chi<1$ this expression is maximized at all values of $0<t<\infty$ by
choosing  $\gamma^u_0= (1+\chi)^{-1}$ (its maximum allowed value) and $\beta^u_0=0$.
From Eqs.~(\ref{shoe})--(\ref{last}) this is achieved only for the unraveling $u=-1$.
For this maximally robust unraveling the survival probability simplifies to
\beq
S^{\cal R}(t) = \sqrt{\frac{1-\chi^2}{1-\chi^2 e^{-(1-\chi)t/2}}}.
\eeq
From Eqs.~(\ref{deftau}) and (\ref{maxev}) the maximum ensemble average 
survival time is thus
\beq
\tau^{\cal R} = \frac{2}{1-\chi}\ln \left[ 4\chi^{2} 
\left(2+\chi^{2}-2\sqrt{1-\chi^{2}}\right)^{-1}  \right].
\eeq
This is plotted as a function of $\chi$ in Fig.~2, along with other parameters of
interest. Note that the survival time grows with $\chi$ from 
$\tau^{\cal R}=2\ln 2$ at
$\chi=0$ and diverges as $(1-\chi)^{-1}$ as $\chi\to 1$. 
This is characteristic of critical slowing down.
The increase in $\tau^{\cal R}$ with $\chi$ does not mean that the ensemble 
$E^{\cal R}$ becomes more robust as $\chi$ increases; on the contrary, for 
fixed $t$, $S^{\cal R}(t)$ decreases monotonically with $\chi$.
 For large $\chi$ the 
survival probability $S^{\cal R}(t)$ decays more quickly, 
but remains greater than the maximum eigenvalue $\Lambda$ (which decreases 
rapidly for large $\chi$) for a longer time.

In Conclusion, maximally robust unravelings are a quant\-it\-at\-ive tool for 
investigating open 
quantum systems, and give insights quite 
distinct from those offered by traditional techniques. Our work 
differs from previous work in that we have considered the  
average survival probability $S(t)$ for {\em an ensemble which is physically 
realizable} from monitoring the environment, rather than the survival 
probability (or change in entropy) of an 
arbitrarily chosen single pure state. Also our
definition of the survival time $\tau$ as {\em the time at which the 
ensemble resulting from monitoring the environment becomes obsolete} is 
fundamentally motivated, unlike other definition such as 
$\tau^{-1} = \frac{d}{dt}S(t)|_{t=0}$ or $S(\tau)=1/e$.
We have shown that for a 
particular class of unravelings, those which are continuous and 
Markovian, an analytical solution is possible for a simple system. 
Our technique is easily generalizable to any 
system which can be linearized, 
although a partly numerical solution may be necessary.

\begin{figure}
	%\centerline{\includegraphics{ }}
	\vspace{6cm}
	\begin{center}
		\special{illustration 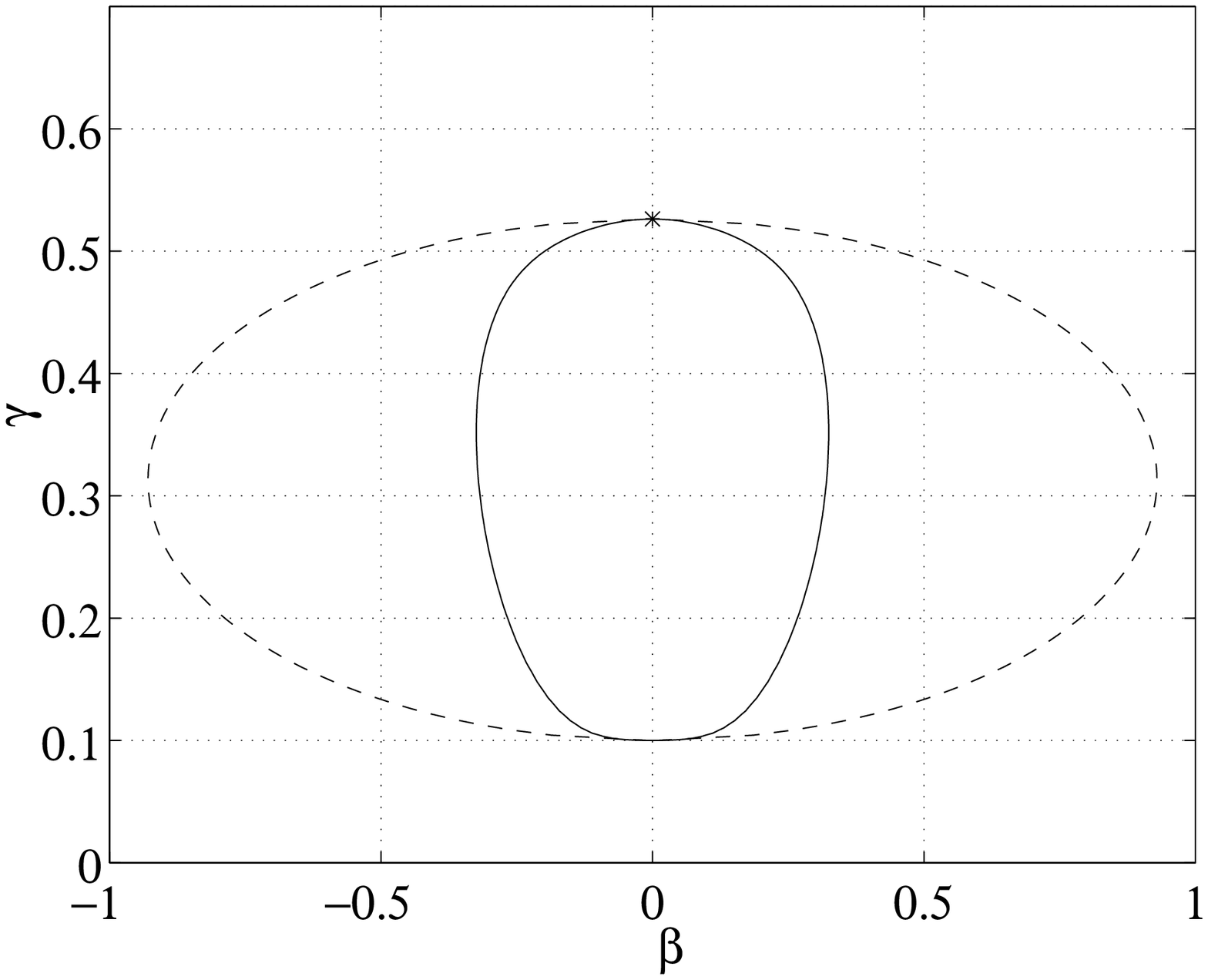 scaled 0.45}
	\end{center}
	\vspace{-0.5cm}
	\caption{\narrowtext Plot of the allowed values of the second-order moments 
	$\gamma=\langle x^{2}\rangle$ and $\beta=\langle 
	xy+yx\rangle/2$ held by all members of an ensemble of Gaussian pure states 
	which represent the stationary state $\rho_{\infty}$. The solid line 
	is the boundary of the allowed region for ensembles $E^{u}$ which are 
	physically realizable through Eq.~(\ref{SSE2}), while the dashed line 
	is for unconstrained ensembles. The star indicates the position of
	 the most robust ensemble $E^{\cal R}$. The threshold parameter $\chi$ 
	equals $0.9$.
	}
	\protect\label{fig1}
\end{figure}

\begin{figure}[h!]
	\vspace{6cm}
	\begin{center}
		\special{illustration 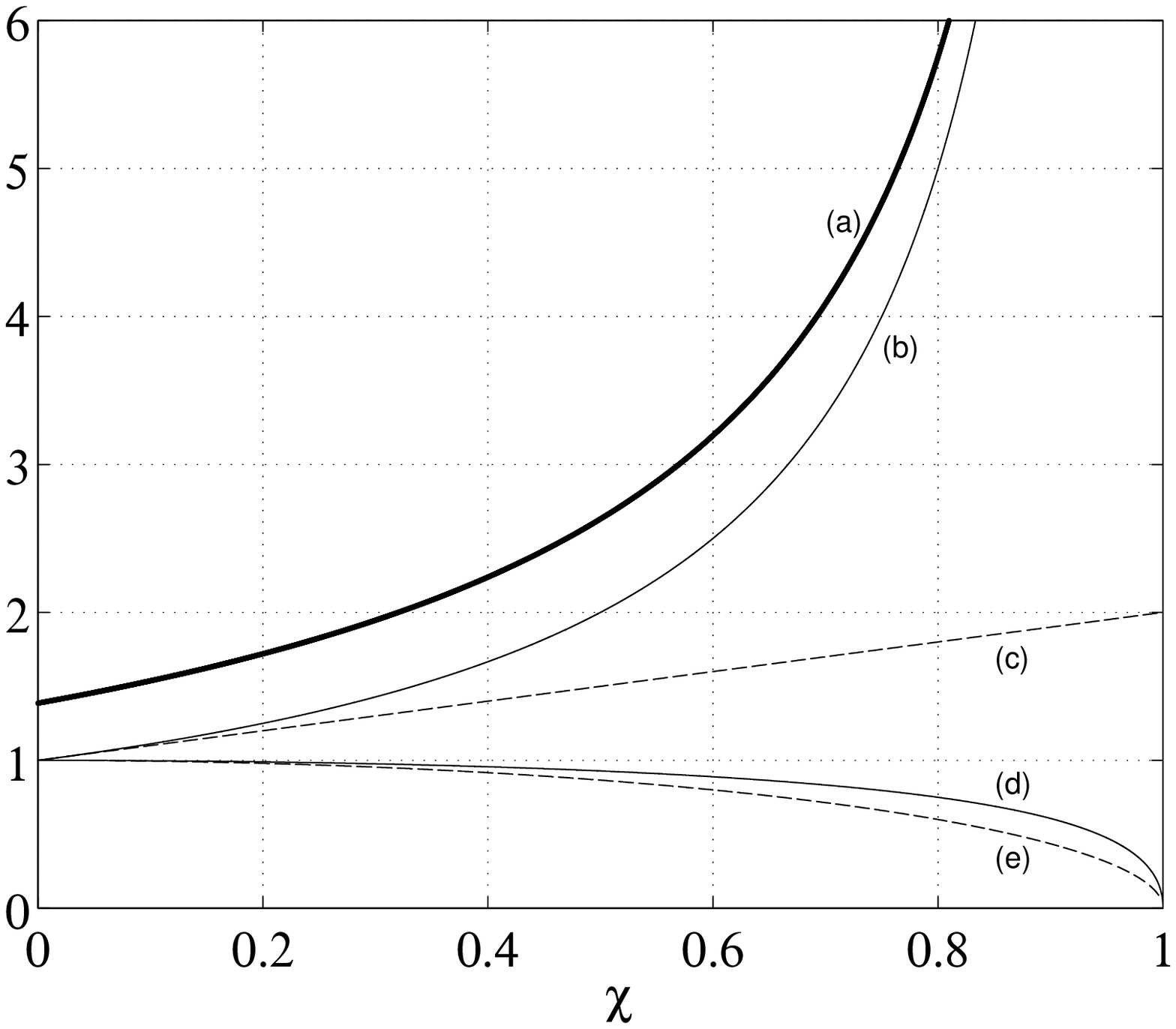 scaled 0.45}
	\end{center}
	\vspace{-0.5cm}
	\caption{\narrowtext Plot of (a) the average survival time $\tau^{\cal 
	R}$ (in units of the decay time) of the maximally robust ensemble $E^{\cal R}$; 
	(b) the $y$-variance $\alpha_{\infty}$
	for the stationary state $\rho_{\infty}$; (c) the $y$-variance 
	$\alpha_{0}^{\cal R}$ for the members of  
	$E^{\cal R}$; (d) the largest eigenvalue 
	$\Lambda$ of $\rho_{\infty}$; and (e) $S(\infty) = {\rm 
	Tr}[\rho_{\infty}^{2}]$ versus threshold parameter $\chi$.  
	}
	\protect\label{fig2}
\end{figure}

\end{multicols}

\end{document}